\documentclass[manuscript]{aastex}
\usepackage[utf8]{inputenc}
\usepackage{graphicx}
\usepackage{amsmath} 
\usepackage{wasysym} 
\usepackage{bm}  

\newcommand{\up}[1]{\ifmmode^{\rm #1}\else$^{\rm #1}$\fi}

\newcommand{\arcd}{\ifmmode^{\circ}\else$^{\circ}$\fi}
\newcommand{\arcm}{\ifmmode{'}\else$'$\fi}
\newcommand{\arcs}{\ifmmode{''}\else$''$\fi}

\slugcomment{\today{} version}

\shorttitle{M31 Double Mode Cepheids}
\shortauthors{Poleski}

\begin{document}


\title{Double-mode Cepheids in M31}

\author{Rados\l{}aw Poleski\altaffilmark{1}}
\email{poleski@astronomy.ohio-state.edu}

\altaffiltext{1}{Department of Astronomy, Ohio State University, 140 W. 18th Ave., Columbus, OH 43210, USA}


\begin{abstract}
Until now, double mode Cepheids (or beat Cepheids) were known in the Galaxy, Magellanic Clouds and M33. 
Curiously, none of more than 2000 Cepheids in M31 was claimed to show two pulsation modes. 
We conducted a systematic search for double mode Cepheids in the archival data and discovered four such objects. 
We identify one of the stars as a first and second overtone pulsator even though its secondary period is subject to strong aliasing.
Two stars turn out to pulsate in the fundamental mode and the first overtone. 
Their fundamental periods are $9.392~{\rm d}$ and $9.163~{\rm d}$.
This makes them the first candidates for the fundamental mode and the first overtone Cepheids, which double mode pulsations are caused by the 1:2 resonance of the fundamental mode and the second overtone. 
\end{abstract}


\section{Introduction} 

Among many different types of pulsating stars, probably the most important group are classical (or type I) Cepheid variables. 
They obey a power-law period-luminosity relation, which allows measurement of distances to external galaxies. 
However, the details of this relation are subject of an active debate. 
The slope of the relation changes around a period of $\approx 10~{\rm d}$ at some wavelengths \citep[e.g.][]{ngeow09,tammann11}. 
Also the zero point of the relation is affected by metallicity. 
Different star formation history in different galaxies 
result in different histograms of Cepheid periods,  
which impacts the distance estimates if the period-luminosity relation is not a pure power-law. 
The other effects that affect the observed brightness of Cepheids are: extinction, infrared emission of circumstellar dust and blending. 
The above problems can not be tackled without a deep understanding of the Cepheid internal structure. 
Only recently has the discrepancy of Cepheid masses derived from the stellar evolution and pulsation theories been solved \citep{pietrzynski10}. 

Among the Cepheid variables an important group constitute multimode radial pulsators. 
The identification of two modes with the expected period ratio clearly shows that 
the object is a classical Cepheid, not a different type of the variable star. 
It also yields an additional constraint on stellar parameters as the two modes probe different parts of the star.
In this context even more important are triple mode pulsators \citep{moskalik05} of which we know a dozen or so. 

The highest number of Cepheids is known in the Magellanic Clouds. 
These galaxies contain altogether more than 8000 Cepheids \citep{soszynski08,marquette09,soszynski10cepsmc,soszynski12}. 
Among them $6\%$ show at least two radial modes in the Small Magellanic Cloud. 
For the more metal-rich Large Magellanic Cloud the corresponding value is $10\%$.
Except these, the double mode Cepheids are known only in the Milky Way and M33 \citep{beaulieu06b}. 
There are also more than 2000 Cepheids known in Andromeda galaxy \citep[M31;][]{fliri12,kodric13} 
and until now none of them has been claimed as a double mode pulsator. 
If the same single mode to double mode number ratio of Cepheids is in the M31 as in Magellanic Clouds one would expect more than a hundred double mode Cepheids to be detected by a deep enough survey. 
The number should be smaller because of more severe blending, which lowers the observed amplitudes of Cepheids and affects most the smaller amplitude mode. 
Even accounting for this, one would expect at least a few longest period (i.e., brightest) double mode Cepheids to be known in M31. 
We note that three double mode RR Lyr variables were found in the Hubble Space Telescope photometry by \citet{reyner10} and they are fainter than Cepheids. 

Here we present a search for double mode Cepheids in archival photometry for known variable stars in M31. 
The next section briefly presents data used in the analysis. 
We present the search method and its results in Sections~3 and 4. 
Finally, we discuss our findings. 
During our research, we found missclassifications for a few Cepheids, which are presented in the Appendix.

\section{Data and method description} 

Our analysis is based on two sets of publicly available lightcurves for the Cepheids in M31.
In their search for  Cepheids in M31 \citet{vilardell07} used data collected between 1999 and 2003 for stars in NE part of this galaxy. 
One field of $0.3~{\rm deg^2}$ was observed with 2.5~m Isaac Newton Telescope. 
The pixel scale was $0.33''$ and median seeing was $1.3''$ in the B band and $1.2''$ in the V band. 
The analysis, based on 265 epochs in B band and 259 in V band led to a discovery of 416 Cepheids.
For more details we refer to \citet{vilardell06,vilardell07}.

The second set of lightcurves was presented by \citet{kodric13}.
They analyzed data collected by $1.8$~m Panoramic Survey Telescope and Rapid Response System (Pan-STARRS)
with $1.4\times10^9~{\rm pixels}$ camera.
The pixel scale is $0.258''$ and a total field of view is $7~{\rm deg^2}$ i.e., it allows covering the whole of M31 in a single image. 
The 183 epochs in $r_{\rm P1}$ and $i_{\rm P1}$ bands \citep{tonry12} collected in 2010 and 2011 were used to search for Cepheids. 
The median seeing of the 30 best seeing images was $0.86''$. 
Altogether 2009 Cepheids were found including type II objects and variables that could not be definitely assigned to type I or type II. 
The more detail description of the observation and their analysis was presented by \citet{lee12} and \citet{kodric13}.
Both \citet{vilardell07} and \citet{kodric13} used  difference image analysis to preform photometric measurements.

The lightcurves of Cepheids were prewhitened with the period given in the original papers 
as well as periods independently found by us. 
The two periods differed only in questionable cases.
The prewhitened lightcurves were searched for periods using both the discrete Fourier transform and
multiharmonic analysis of variance \citep{schwarzenbergczerny96}. 
The final period estimates were taken using the latter method results.
All the secondary periods found in a wide range around known structures in the Petersen diagram (period ratio vs. longer period)
were visually verified. 
During our analysis we found a few stars that were incorrectly classified. 
We present these in the Appendix in order to help purify the M31 Cepheid sample used in future studies.

\section{Results} 

Below we present each double mode pulsator separately. 
Comparison to the Cepheids found in other environments and discussion of pulsation properties will be presented in next section. 
We denote fundamental mode and first overtone pulsators by F/1O and first and second mode pulsators by 1O/2O. 
The star identifiers come from \citet{vilardell07} and \citet{kodric13}.
Periods and Fourier parameters \citep[phase differences $\phi_{21}=\phi_2-2\phi_1$ and amplitude ratios $R_{21}=A_2/A_1$;][]{simon81}
are presented in Table~\ref{tab:basic} for each mode separately.

\begin{deluxetable}{lrrrrrr}
\tabletypesize{\scriptsize}
\tablecaption{Photometric properties of double mode M31 Cepheids.\label{tab:basic}}
\tablewidth{0pt}
\tablehead{
\colhead{ID} & 
\colhead{$P_1~[{\rm d}]$} & \colhead{$\phi_{21,1}$} & \colhead{$R_{21,1}$} &
\colhead{$P_2~[{\rm d}]$} & \colhead{$\phi_{21,2}$} & \colhead{$R_{21,2}$} 
}
\startdata
J00450019+4129313      & $1.694919(16)$ & $4.13$ & $0.17$ & $1.361279(27)$ & $2.74$ & $0.12$ \\ 
PSO\_J010.6063+40.8608 & $9.3918(85)$   & $4.99$ & $0.09$ & $6.5551(49)$   & $4.11$ & $0.18$ \\ 
PSO\_J010.9364+41.2504 & $9.1633(80)$   & $4.79$ & $0.17$ & $6.3618(61)$   & $5.32$ & $0.07$ \\ 
PSO\_J011.3583+42.0404 & $10.4672(74)$  & $3.89$ & $0.10$ & $6.1610(52)$ & $4.41$ & $0.49$ \\ 
\enddata
\end{deluxetable}

\subsection{J00450019+4129313 -- 1O/2O type pulsator} 

\begin{figure}
\epsscale{.72}
\plotone{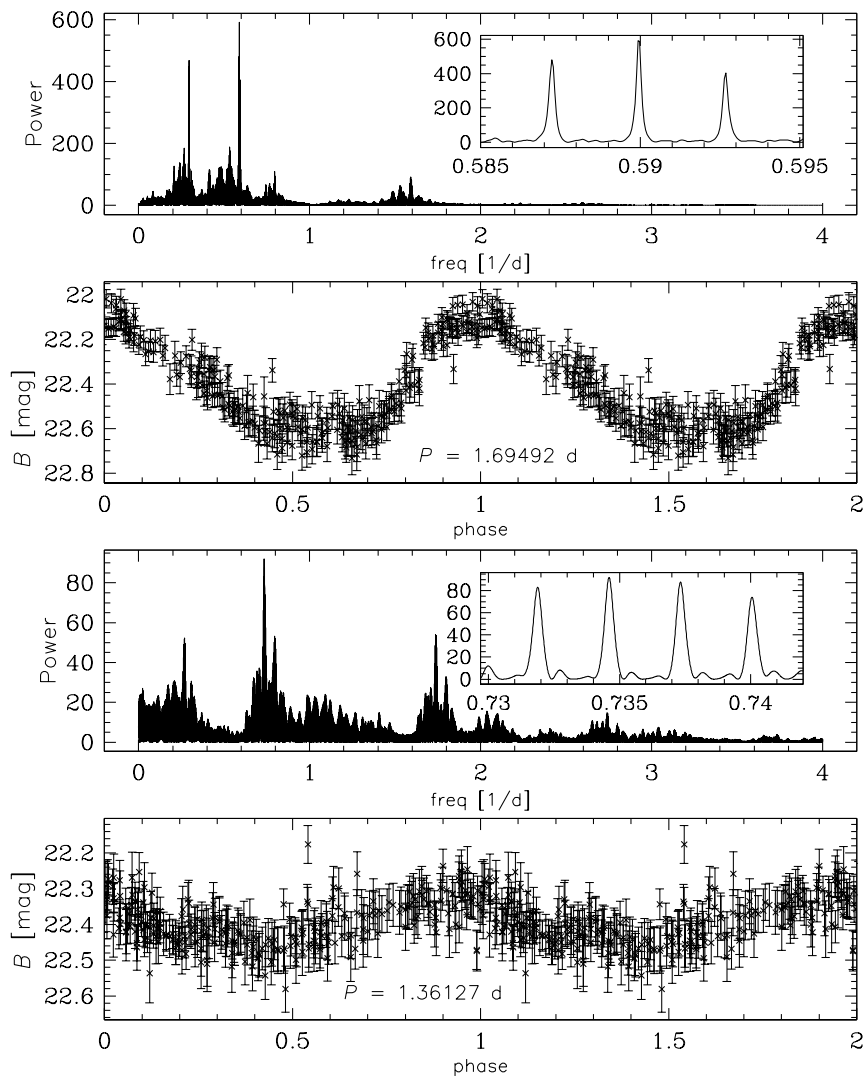}
\caption{Prewhitening of J00450019+4129313 photometry. 
The top panel presents the periodogram of the original data.
The second panel shows the data phased with the period of highest peak of the periodogram.
After subtracting the Fourier fit we recalculated the periodogram (third panel) and phased the data according to the highest peak (bottom panel). 
The insets in the first and the third panels magnify the parts of periodogram close to the highest peak. 
\label{fig:cep257}}
\end{figure}

The photometry presented by \citet{vilardell07} reveals the primary period of this star to be $1.694919(16)~{\rm d}$. 
We found a strong signal for a secondary period but its value cannot be found unambiguously 
because of the strong aliases in the power spectrum. 
The prewhitening of the light curve is illustrated in Figure~\ref{fig:cep257}. 
The four possible values are (starting from the most probable): 
$1.361279(27)$,
$1.356241(44)$,
$1.366328(47)$, or
$1.351260(51)$.
They give period ratios of
$0.8032$,
$0.8002$,
$0.8061$, and
$0.7972$, respectively.
All of these values fall in the range typical for 1O/2O pulsators.

\subsection{PSO\_J010.6063+40.8608 -- F/1O type pulsator} 

\begin{figure}
\epsscale{.72}
\plotone{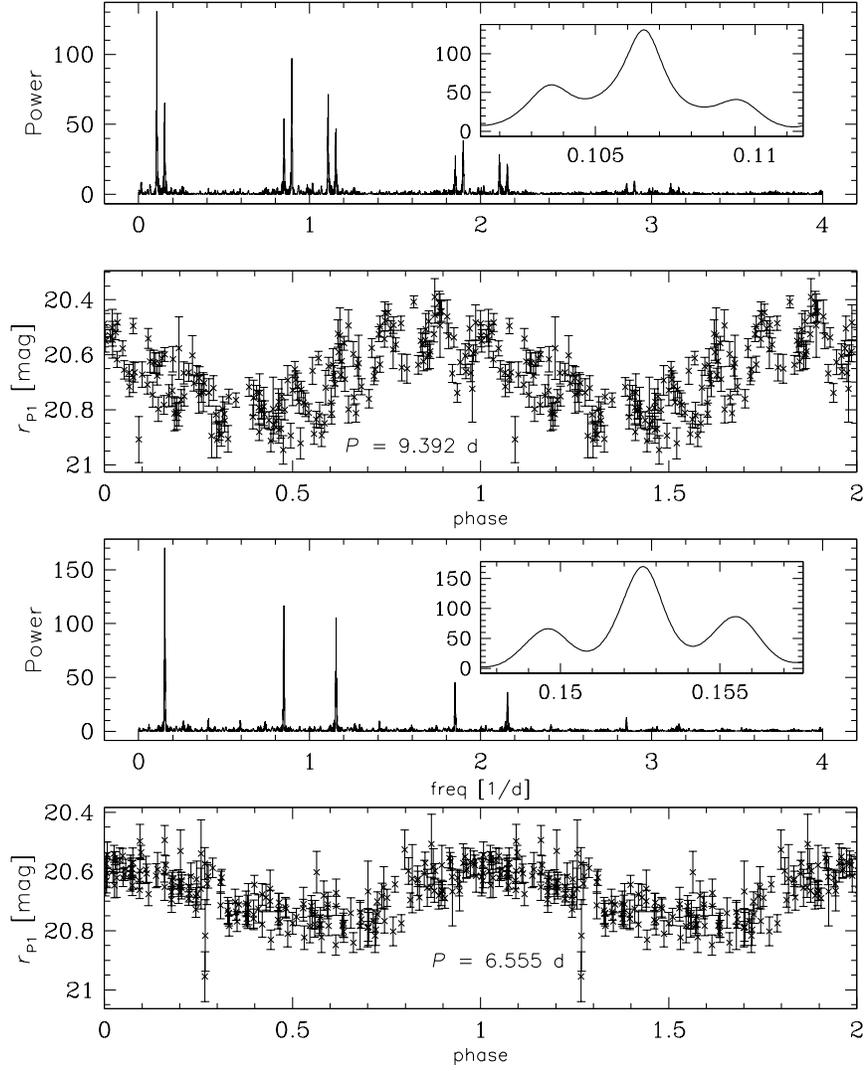}
\caption{Same as Figure~\ref{fig:cep257} for PSO\_J010.6063+40.8608.
\label{fig:cep1300}}
\end{figure}

\citet{an04} found the period of this star to be $9.42~{\rm d}$. 
The photometry presented by \cite{kodric13} results in a primary period of $9.3918(85)~{\rm d}$ 
and a secondary period of $6.5551(49)~{\rm d}$. 
The prewhitening of the light curve is presented in Figure~\ref{fig:cep1300}. 
The period ratio of $0.698$ is slightly lower than the period ratio for known F/1O pulsators but all of them have shorter periods. 
It lies on the extension of F/1O sequence on the Petersen diagram.
Thus we classify it as F/1O Cepheid.

\subsection{PSO\_J010.9364+41.2504 -- F/1O type pulsator} 

\begin{figure}
\epsscale{.72}
\plotone{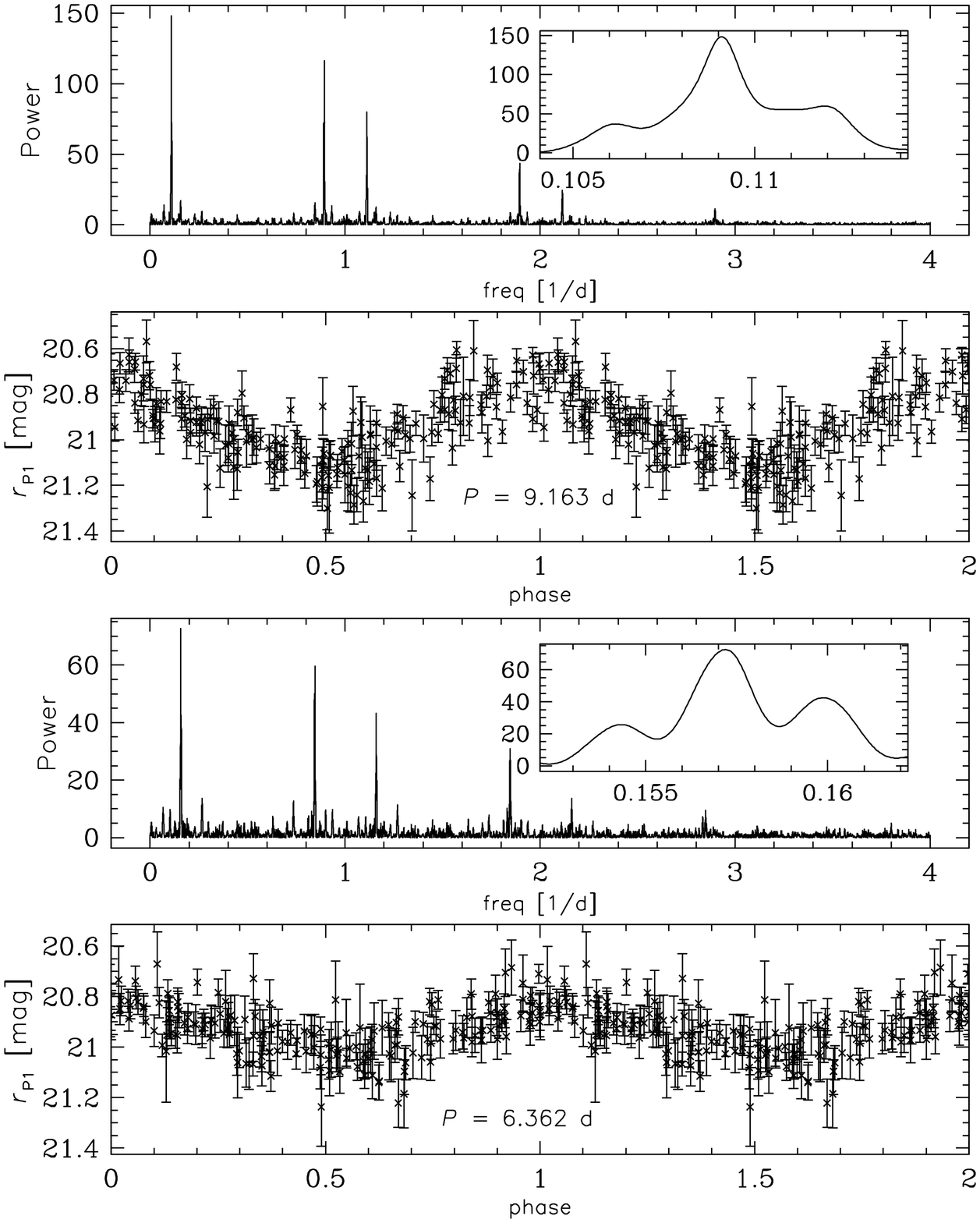}
\caption{Same as Figure~\ref{fig:cep257} for PSO\_J010.9364+41.2504.
\label{fig:cep1285}}
\end{figure}

This object was already announced by \citet{kaluzny99} and \citet{joshi03} with periods of $9.173~{\rm d}$ and $9.160~{\rm d}$. 
Using \citet{kodric13} photometry we found periods of $9.1633(80)~{\rm d}$ and $6.3618(61)~{\rm d}$ (Figure~\ref{fig:cep1285}). 
This object turns out to have periods similar to PSO\_J010.6063+40.8608 discussed above, 
with a slightly smaller period ratio of $0.694$.
We conclude that this is a F/1O pulsator. 

\subsection{PSO\_J011.3583+42.0404 -- candidate non-radial pulsator} 

\begin{figure}
\epsscale{.72}
\plotone{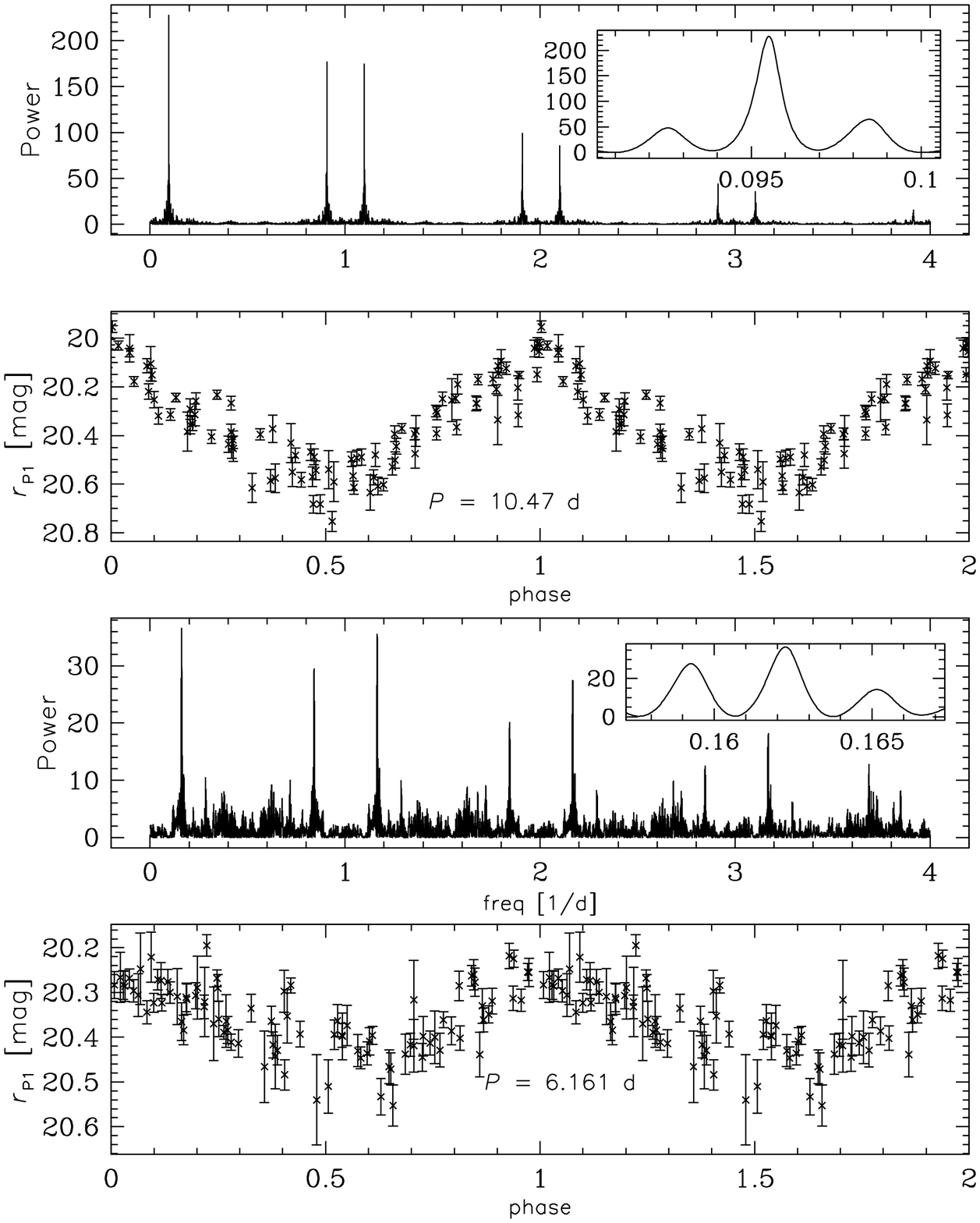}
\caption{Same as Figure~\ref{fig:cep257} for PSO\_J011.3583+42.0404.
\label{fig:cep1364}}
\end{figure}

The analysis of photometry presented by \citet{kodric13} revealed two periods in this object:
$10.4672(74)~{\rm d}$ and
$6.1610(52)~{\rm d}$. 
The prewhitening shown in Figure~\ref{fig:cep1364} reveals a sound detection of secondary period. 
The period ratio of $0.589$ is not typical for any known combination of radial modes. 
This object can be either a non-radial pulsator or a blend of two stars.
Non-radial modes are observed in other Cepheids \citep{moskalik09}.
We note that \citet{dziembowski12} tried to reproduce $\approx0.6$ period ratio observed in some of the LMC and SMC first overtone Cepheids. 
Their findings are not applicable to PSO\_J011.3583+42.0404 because neither analyzed sample nor the models extended to primary periods long enough. 

\section{Discussion} 

\begin{figure}
\epsscale{.9}
\plotone{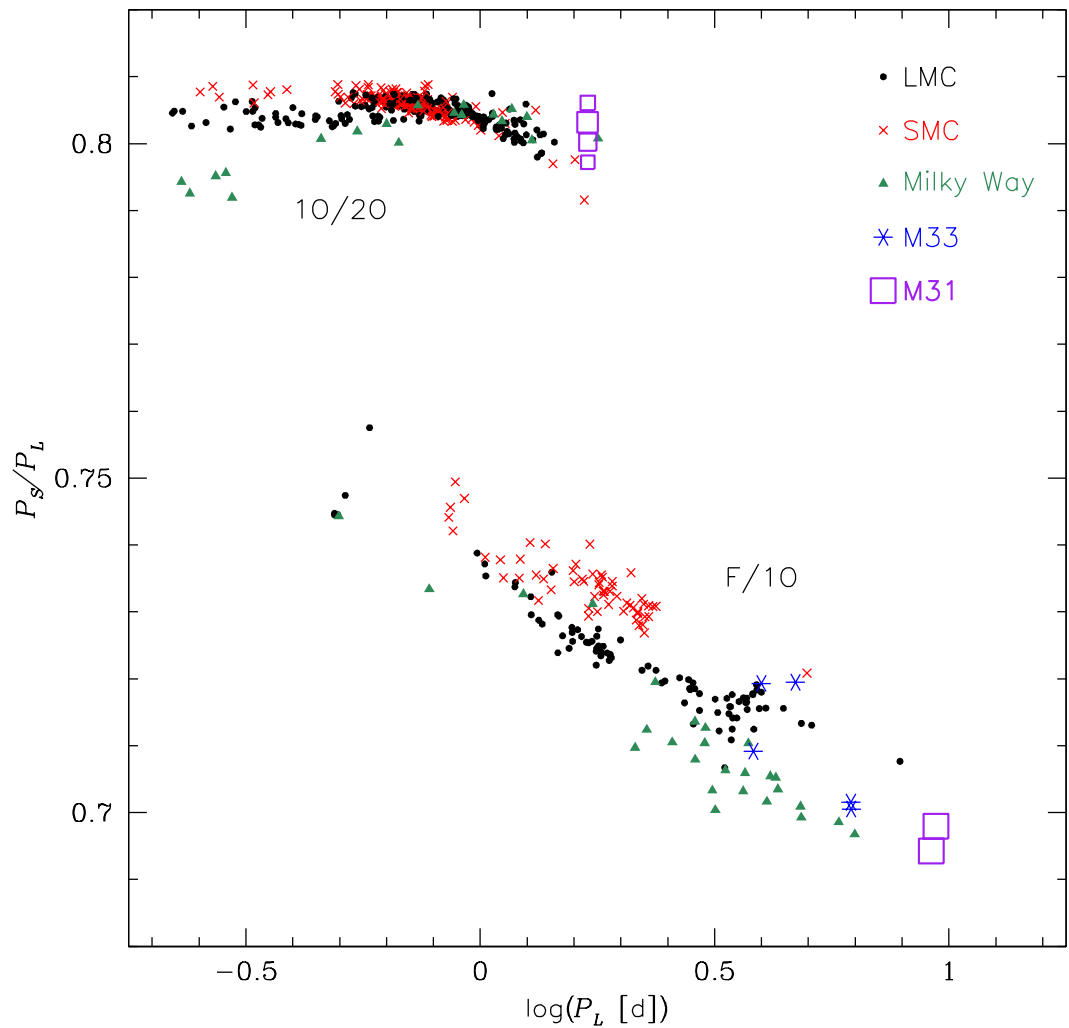}
\caption{Petersen diagram for known F/1O and 1O/2O Cepheids. 
The number of Cepheids plotted is 355, 281, 49, 5 and 3 for LMC, SMC, Milky Way, M33 and M31, respectively. 
Four possible positions of M31 1O/2O Cepheid J00450019+4129313 are presented with larger symbols corresponding to the more probable positions. 
\label{fig:peter}}
\end{figure}

We show the Petersen diagram for known classical double mode F/1O and 1O/2O Cepheids in Figure~\ref{fig:peter}. 
Separate symbols are used to present objects from the Milky Way \citep[][and reference therein]{soszynski11,smolec10}, LMC \citep{soszynski08,marquette09,soszynski12}, SMC \citep{soszynski10cepsmc}, M33  \citep{beaulieu06b} and M31. 
There are four different positions shown for J00450019+4129313, which correspond to the different aliases of the secondary period. 
All are consistent with 1O/2O pulsator. 
One can see that positions of PSO\_J010.6063+40.8608 and PSO\_J010.9364+41.2504 are consistent with extrapolation of the relation seen for F/1O pulsators. 
The fact that fundamental mode periods of these objects are close to $10~{\rm d}$ makes them unusual. 
The lightcurves of the fundamental mode Cepheids with similar periods show two maxima in each period.
The appearance of the second maximum is caused by a 1:2 resonance of the second overtone and the fundamental mode. 
We note that the calculations of \citet{buchler09}, which were presented only in conference proceedings, 
suggest that F/1O pulsations with fundamental mode periods of around $10~{\rm d}$ are caused by the resonance mentioned above. 
The range of luminosities and effective temperatures in which this mechanism operates is very small. 
This should allow very detail modeling of these stars. 
Also the preliminary models for these stars indicate $P_{\rm F}/P_{\rm 2O}$ close to 2 and masses from 6 to $7~{\rm M_{\odot}}$ (Dziembowski \& Smolec, in preparation). 
Based on the Petersen diagram presented in~Figure~\ref{fig:peter} we suggest that the longest period F/1O Cepheids in LMC (OGLE-LMC-CEP-1082, $P_{\rm F}=7.86434~{\rm d}$ and $P_{\rm 1O}=5.56518~{\rm d}$) and SMC (OGLE-SMC-CEP-1497, $P_{\rm F}=4.9780~{\rm d}$ and $P_{\rm 1O}=3.588329~{\rm d}$) may also display double mode pulsations because of the 1:2 resonance of 2O and F modes. 
Both these objects are separated from the rest of F/1O Cepheids in a given galaxy by at least $2.3~{\rm d}$ in $P_{\rm F}$.

The double-mode Cepheids with periods as long as presented here might have been overlooked in previous analyses of the existing time-series photometry for variable stars in other galaxies. 
The double mode pulsators presented here fall outside not only the previously published Petersen diagrams constructed for known pulsators but also outside the period range studied in many theoretical investigations. 

Finally, we would like to comment on a few different data access policies that we have seen during the literature search for data on the analyzed stars. 
One paper discussed the results of the search for M31 Cepheids and scientific results found using them but did not provide any information about the stars themselves, not even their coordinates. 
The next one presented the coordinates and basic properties of the variable stars but did not presented any time-resolved photometry. 
In other example the authors presented the list of variables and claimed that the timeseries photometry is made public but in fact it was not accessible. 
In one more example we have found that the photometry is published but the Julian Dates are rounded to integer values, which makes such photometry useless. 
There were only two examples in which the authors made photometry public in a proper form to allow scientific research.  
We call on authors to present both the properties of the variable stars found (at least position, brightness, and period) as well as their photometric timeseries for the astronomical community.
With growing observing capabilities, observers typically are not able to extract the whole scientific information contained in their data.
Allowing others to conduct more detail investigation is a good practice.

\acknowledgments
The author is grateful Wojciech Dziembowski and Rados\l{}aw Smolec for fruitful discussion as well as Andy Gould for reading the manuscript. 
This research has made use of the VizieR catalogue access tool, CDS, Strasbourg, France.

\appendix
\section{Misclassified objects} 

\paragraph{J00451769+4136367} 

\citet{vilardell07} gives period of $7.862587~{\rm d}$. 
The true period is two times shorter.

\paragraph{J00452829+4139547} 

Photometry presented by \citet{vilardell07} indicates that this is an artifact produced by the difference image analysis.
The variations of measured flux are caused by the nearby brighter Cepheid J00452353+4140040. 
The periods of both objects agree.

\paragraph{PSO\_J009.9994+40.5558} 

Our analysis of photometry presented by \citet{kodric13} revealed that this object varies with periods of $13.084(14)~{\rm d}$ and $9.5589(80)~{\rm d}$. 
The period ratio of $0.7306$ is similar to F/1O pulsators. 
However, this object cannot be classified as a double mode pulsator. 
As noted by \citet{baade65} there are two Cepheids of similar brightness and separated by around $1.8''$. 
\citet{baade65} identified them by numbers of 252 and 253.
Their periods agree with quoted above. 
We conclude that \citet{kodric13} presented  photometry of two blended stars. 

\paragraph{PSO\_J010.5263+40.7724} 

Classification of this object by \citet{kodric13} as a Cepheid seems questionable as its light curve does not resemble typical Cepheids. 
The $i_{\rm P1}$-band photometry does not show clear periodic variability.

\bibliographystyle{apj}

\end{document}